\documentclass[aps,prc,letterpaper,11pt,twoside,tightenlines,nofootinbib,showpacs,preprint,superscriptaddress]{revtex4}
\usepackage{amsmath}
\usepackage{graphicx}
\usepackage{subfigure}
\usepackage{color}
\RequirePackage[colorlinks,citecolor=blue,urlcolor=blue,linkcolor=blue]{hyperref}

\newcommand{\mev}{\textrm{ MeV}}

\begin{document}

\title{$J/\psi N$ interactions revisited and $\Lambda_b^0 \to J/\psi K^- (\pi^-) p$ decays}

\author{C. W. Xiao}
\affiliation{Institut  f\"{u}r Kernphysik (Theorie), Institute for Advanced Simulation, and J\"ulich Center for Hadron Physics, Forschungszentrum J\"ulich, D-52425 J\"{u}lich, Germany}

\begin{abstract}
Recently LHCb experiments confirmed the findings of the two $P_c$ states in the $\Lambda_b^0 \to J/\psi p \pi^-$ decays. In the present work, we investigate both the $\Lambda_b^0 \to J/\psi K^- p$ and the $\Lambda_b^0 \to J/\psi p \pi^-$ decays, continuing the investigations of our former works on the interactions of $J/\psi N$ with its coupled channels by considering the s-/u- channel contributions. We obtain consistent results of the line-shape of the $J/\psi N$ invariant mass distribution with the LHCb experiments, and favour the $P_c (4450)$ state as a $\bar{D}^* \Sigma_c$ bound state with $J=1/2^-$. 

\end{abstract}

\pacs{}
\maketitle
\date{}

\section{Introduction}

Since $X(3872)$ was found by Belle in 2003 \cite{Choi:2003ue}, much effort has been made to understand the properties of the ``exotic" states and to search new states both in the theoretical and experimental aspects. Such ``exotic'' states can not be understood as the normal color-singlet hadrons with the structure of quark-antiquark (mesons) or three quarks (baryons), which could be multi-quark structures, for example tetraquark and pentaquark states (more discussions can be found in the reviews \cite{Chen:2016qju,Hosaka:2016pey}). Recently, two new $P_c^+$ states were found by LHCb in the $J/\psi p$ mass spectrum of the $\Lambda_b^0 \to J/\psi K^- p$ decays \cite{Aaij:2015tga} using a model-dependent analysis, which are considered as charmonium-pentaquark states with some uncertainties about their spin-parity $J^P$ quantum numbers and were predicted in the early works \cite{Wu:2010jy,Wu:2010vk,Yang:2011wz,Wu:2012md,Garcia-Recio:2013gaa,Xiao:2013yca} and confirmed by a model-independent re-analysis with the previous experimental data sample \cite{Aaij:2016phn}. These findings are mostly explained as molecular states \cite{Chen:2015loa,Chen:2015moa,Roca:2015dva,He:2015cea,Meissner:2015mza,Lu:2016nnt,Shimizu:2016rrd,Shen:2016tzq,Ortega:2016syt,Yamaguchi:2016ote}, even though Ref. \cite{Mironov:2015ica} questions about their molecular properties. On the other hand, with different theoretical models, they also can be explained as a diquark state \cite{Maiani:2015vwa,Lebed:2015tna,Anisovich:2015cia,Li:2015gta,Wang:2015epa,Wang:2015ava,Zhu:2015bba,Wang:2015ixb,Ali:2016dkf}, a compact pentaquark state \cite{Santopinto:2016pkp}, a soliton-$\bar{D}$-$D$ bound state \cite{Scoccola:2015nia}, or contrarily a kinematical effect or a cusp effect \cite{Guo:2015umn,Liu:2015fea,Mikhasenko:2015vca}. As suggested in Refs. \cite{Burns:2015dwa,Wang:2015pcn}, LHCb found a new evidence of the observation about these two states by a full amplitude analysis of the $\Lambda_b^0 \to J/\psi p \pi^-$ decays \cite{Aaij:2016ymb}. It was remarked before in Ref. \cite{Guo:2016bkl} that the similar triangle singularity will also emerge in the $J/\psi p$ invariant mass distribution as in the $\Lambda_b^0 \to J/\psi K^- p$ decays. On the other hand, the role of $\Lambda (1405)$ in the $\Lambda_b^0 \to J/\psi K^- p$ decays was discussed with both Murcia-Valencia model and Bonn model in Ref. \cite{Roca:2015tea} before the experiments. Thus, about the existence of these two pentaquark states, the confidence and the suspicion coexist since there are different theoretical opinions on them. Based on the former investigations, in the present work we go further to study the two $P_c$ states theoretically.

Using the chiral unitary approach, where only the leading-order contact terms (Weinberg-Tomozawa, WT) from the lowest order Lagrangian are taken into account, Ref. \cite{Oset:1997it} dynamically reproduces the $\Lambda (1405)$ resonance in the coupled channel $\bar{K} N$ s-wave interactions. With the same approach, but considering the s- and u- channel diagrams (direct and crossed graphs, or Born terms, as they called), the $\bar{K} N$ interactions are revisited in Ref. \cite{Oller:2000fj} where the experimental data are described well. One step further, up to next-to-leading order contributions from the Lagrangian, also including the s- and u- channel diagrams of the leading order contributions, the $\bar{K} N$ interactions and the $\Lambda (1405)$ resonance are investigated again in Refs. \cite{Borasoy:2005ie,Oller:2005ig} with the analysis of the new experimental data. Along the same line, the improved analysis are done in Refs. \cite{Borasoy:2006sr,Ikeda:2012au,Hyodo:2011ur,Mai:2012dt,Guo:2012vv} and in recent works \cite{Mai:2014xna,Cieply:2016jby,Ramos:2016odk} considering on-shell or off-shell contributions in the interaction potentials. Within a framework based on the chiral and hidden local symmetries, Ref. \cite{Khemchandani:2011mf} finds a strong coupling of the vector meson-baryon (VB) channels \footnote{The approximations made in this formalism are being scrutinized~\cite{GMMO}.} to the resonances $\Lambda (1405)$, $\Lambda (1670)$ which are dominated by the pseudoscalar meson-baryon (PB) dynamics. Further, the contributions of the s-/u- channel diagrams, and the Kroll-Ruderman terms in the PB $\to$ VB transitions are taken into account in the later work \cite{Khemchandani:2012ur}. Our former papers \cite{Xiao:2013yca,Xiao:2015fia} only consider the coupling of the VB channels to the PB channels and without taking onto account the contributions from the s- and u- channel diagrams. This is the motivation of the present work. In principle, one could do up to next to leading order calculations. But, since the lack of experimental data in the heavy quark sector which is not like the case of the $\bar{K} N$ interactions, Ref. \cite{Lu:2014ina} faces with such problems for determining the free parameters at next to leading order. Therefore, in the present work, we do not take into account the next-to leading order contributions. In the next section, we present our modified formalism. Then, we discuss the contributions of the s- and u- channel diagrams in the third section. Next, we investigate the the $\Lambda_b^0 \to J/\psi K^- (\pi^-) p$ decays with the obtained $J/\psi p$ amplitudes. Finally, we finish with our conclusions.

\section{$J/\psi N$ interactions}

In the early work \cite{Xiao:2013yca}, using the local hidden gauge formalism \cite{Meissner:1987ge,Bando:1987br,Harada:2003jx} and combining the heavy quark spin symmetry (HQSS) \cite{Neubert:1993mb,hqss00}, we have considered seven coupled channels with $J/\psi N$: $\eta_c N$, $\bar D \Lambda_c$, $\bar D \Sigma_c$, $\bar D^* \Lambda_c$, $\bar D^* \Sigma_c$, $\bar D \Sigma_c^*$, $\bar D^* \Sigma^*_c$ which can be specified with spin $J$ and isospin $I$ in different sectors. In the present work, we only focus on two sectors as in Ref. \cite{Xiao:2015fia} where the resonances appear. The elements of the interaction potential kernel $V_{ij}$ are given in Tables~\ref{tab:vij11} and \ref{tab:vij31} for the $J=1/2, \, I=1/2$ and $J=3/2, \, I=1/2$ sectors respectively, where the coefficients $\mu_{i}^I$, $\mu_{ij}^I$ ($i,j=1,2,3$) and $\lambda_2^I$ are the unknown low energy constants in the HQSS formalism, which specify the isospin sector and can be related to each other using $SU(3)$ flavour symmetry. As a consequence of the HQSS constraints, all of them just depend on the isospin ($I$) and are independent of the spin $J$ (for more details see Ref.  \cite{Xiao:2013yca}). The values of them depend on the considered model, where we use the local hidden gauge formalism which is different compared with the one in Ref. \cite{Garcia-Recio:2013gaa}. Then, their values for the two considered sectors are given by
\begin{equation}
\begin{split}
\mu_2 &= \frac{1}{4f^2}\;  N_i\;N_j (2\sqrt{s} - M_i - M_j),\\
\mu_3 &= -\frac{1}{4f^2}\;  N_i\;N_j (2\sqrt{s} - M_i - M_j),\\
\mu_{12} &= -\sqrt{6}\ \frac{m_\rho^2}{p^2_{D^*} - m^2_{D^*}}\; \frac{1}{4f^2}\ N_i\;N_j (2\sqrt{s} - M_i - M_j),\\
\mu_1 &= 0,\qquad \mu_{23} = 0, \\
\lambda_2 &= \mu_3,\qquad \mu_{13} = -\mu_{12}~,
\end{split}
\label{eq:vpoten}
\end{equation}
where $s$ is the Mandelstam variable of the meson and the baryon system, $p_{D^*}$ and $m_{D^*}$ are the four momentum and the mass of $D^*$, $f$ the pion decay constant, and the normalization factor for the baryonic fields $N_i = \sqrt{\frac{M_i + E_i}{2 M_i}}$ with $M_i$ and $E_i$ the masses and the energies of the corresponding baryons in the $i^{th}$ channel. For the channels related with the vector mesons, we have ignored the factor $\vec{\epsilon}_i \cdot \vec{\epsilon}_j$. Note that Eqs. \eqref{eq:vpoten} are a bit different from the ones used in Refs. \cite{Xiao:2013yca,Xiao:2015fia} where we have mainly ignored the normalization factors $N_{i (j)}$ and replaced the $k_i^0 + k_j^0$ term by $(2\sqrt{s} - M_i - M_j)$, where $k_i^0, \, k_j^0$ are the energies of incoming, outgoing mesons. Indeed, $N_{i (j)} \approx 1$ in the low energy region as in our cases, we recover these factors for the consistency of the later calculations.

\begin{table}[ht]
     \renewcommand{\arraystretch}{1.7}
     \setlength{\tabcolsep}{0.4cm}
\centering
\caption{The elements  $V_{ij}$ corresponding to the channels in the  $J=1/2,~I=1/2$ sector.}
\label{tab:vij11}
\begin{tabular}{ccccccc}
\hline\hline
$\eta_c N$ & $J/\psi N$ &  $\bar D \Lambda_c$ &  $\bar D \Sigma_c$ &  $\bar D^* \Lambda_c$
  &  $\bar D^* \Sigma_c$ &  $\bar D^* \Sigma^*_c$   \\
\hline
$\mu_1$ & 0 & $\frac{\mu_{12}}{2}$ &
 $\frac{\mu_{13}}{2}$ & $\frac{\sqrt{3} \mu_{12}}{2}$ &
 $-\frac{\mu_{13}}{2 \sqrt{3}}$ & $\sqrt{\frac{2}{3}} \mu_{13}$ \\
  & $\mu_1$ & $\frac{\sqrt{3} \mu_{12}}{2}$ & $-\frac{\mu_{13}}{2 \sqrt{3}}$ & $-\frac{\mu_{12}}{2}$
 & $\frac{5 \mu_{13}}{6}$ & $\frac{\sqrt{2}\mu_{13}}{3}$ \\
  &  & $\mu_2$ & 0 & 0 & $\frac{\mu_{23}}{\sqrt{3}}$ & $\sqrt{\frac{2}{3}} \mu_{23}$ \\ 
  &  &  & $\frac{1}{3} (2 \lambda_2 + \mu_3)$ & $\frac{\mu_{23}}{\sqrt{3}}$ & $\frac{2 (\lambda_2 - \mu_3)}{3 \sqrt{3}}$ & $\frac{1}{3} \sqrt{\frac{2}{3}} (\mu_3-\lambda_2 )$ \\ 
  &  &  &  & $\mu_2$ & $-\frac{2 \mu_{23}}{3}$ & $\frac{\sqrt{2} \mu_{23}}{3}$ \\ 
  &  &  &  &  & $\frac{1}{9} (2 \lambda_2 +7 \mu_3)$ & $\frac{1}{9} \sqrt{2} (\mu_3-\lambda_2)$ \\ 
  &  &  &  &  &  & $\frac{1}{9} (\lambda_2+8 \mu_3)$ \\
\hline
\end{tabular}
\end{table}

\begin{table}[ht]
     \renewcommand{\arraystretch}{1.7}
     \setlength{\tabcolsep}{0.4cm}
\centering
\caption{The elements  $V_{ij}$ corresponding to  the channels in the $J=3/2,~I=1/2$ sector.}
\label{tab:vij31}
\begin{tabular}{ccccc}
\hline\hline
$J/\psi N$ &  $\bar D^* \Lambda_c$ &  $\bar D^* \Sigma_c$ 
&  $\bar D \Sigma^*_c$  &  $\bar D^* \Sigma^*_c$   \\
\hline
$\mu_1$ & $\mu_{12}$ & $\frac{\mu_{13}}{3}$ & $-\frac{\mu_{13}}{\sqrt{3}}$ & $\frac{\sqrt{5} \mu_{13}}{3}$ \\
  & $\mu_2$ & $\frac{\mu_{23}}{3}$ & $-\frac{\mu_{23}}{\sqrt{3}}$ & $\frac{\sqrt{5} \mu_{23}}{3}$ \\
  &  & $\frac{1}{9} (8 \lambda_2 + \mu_3)$ & $\frac{\lambda_2 - \mu_3}{3 \sqrt{3}}$ & $\frac{1}{9} \sqrt{5} (\mu_3-\lambda_2)$  \\ 
  &  &  & $\frac{1}{3} (2 \lambda_2 +\mu_3)$ &
   $\frac{1}{3} \sqrt{\frac{5}{3}} (\lambda_2 -\mu_3)$  \\ 
  &  &  &  & $\frac{1}{9} (4 \lambda_2 +5 \mu_3)$  \\ 
\hline
\end{tabular}
\end{table}

With the chiral unitary approach, the scattering amplitudes are evaluated by solving the coupled channels Bethe-Salpeter equation using the on-shell factorization \cite{Oset:1997it,Oller:2000fj}
\begin{equation}
T = [1 - V \, G]^{-1}\, V,
\label{eq:BS}
\end{equation}
where the propagator $G$ is a diagonal matrix with the meson-baryon loop functions, with the dimensional regularization \footnote{A general expression for n-dimensions can be found e.g. in Ref. \cite{Djukanovic:2009gt}.} the elements of which are given by
\begin{equation}
\begin{split}
G_{ii} (s) =& \frac{2 M_i}{16 \pi^2} \Big\{ a_\mu + \ln \frac{M_i^2}{\mu^2} + \frac{m_i^2 - M_i^2 +s}{2s} \ln \frac{m_i^2}{M_i^2}  \\
& + \frac{q^{cm}_i}{\sqrt{s}} \big[ \ln(s - (M_i^2 - m_i^2) + 2 q^{cm}_i \sqrt{s})  \\
& + \ln(s + (M_i^2 - m_i^2) + 2 q^{cm}_i \sqrt{s})  \\
& - \ln(-s - (M_i^2 - m_i^2) + 2 q^{cm}_i \sqrt{s})  \\
& - \ln(-s + (M_i^2 - m_i^2) + 2 q^{cm}_i \sqrt{s}) \big] \Big\},  \\
\end{split}
\label{eq:giidr}
\end{equation}
where $m_i, ~M_i$ are the masses of meson and baryon in $i^{\rm th}$ channel, respectively, and $q^{cm}_i$ is the three-momentum in the center of mass frame. The only free parameters are $a_\mu$ and $\mu$, taking $a_\mu = -2.3$ and $\mu = 1000 \mev$ as done in Refs. \cite{Wu:2010jy,Wu:2010vk}, which are within the natural values  \cite{Oller:2000fj}. The kernel matrix  $V$ contains the interaction potentials which are discussed above. One should keep in mind that, in our present work we use the chiral unitary approach for the s-wave projections of the interaction potentials, but, for the higher partial waves interactions the off-shell contributions should be considered as discussed in Ref. \cite{Borasoy:2007ku} where an approach is constructed for the meson photoproduction based on the chiral effective Lagrangian.

\begin{figure}
\centering
\includegraphics[scale=0.6]{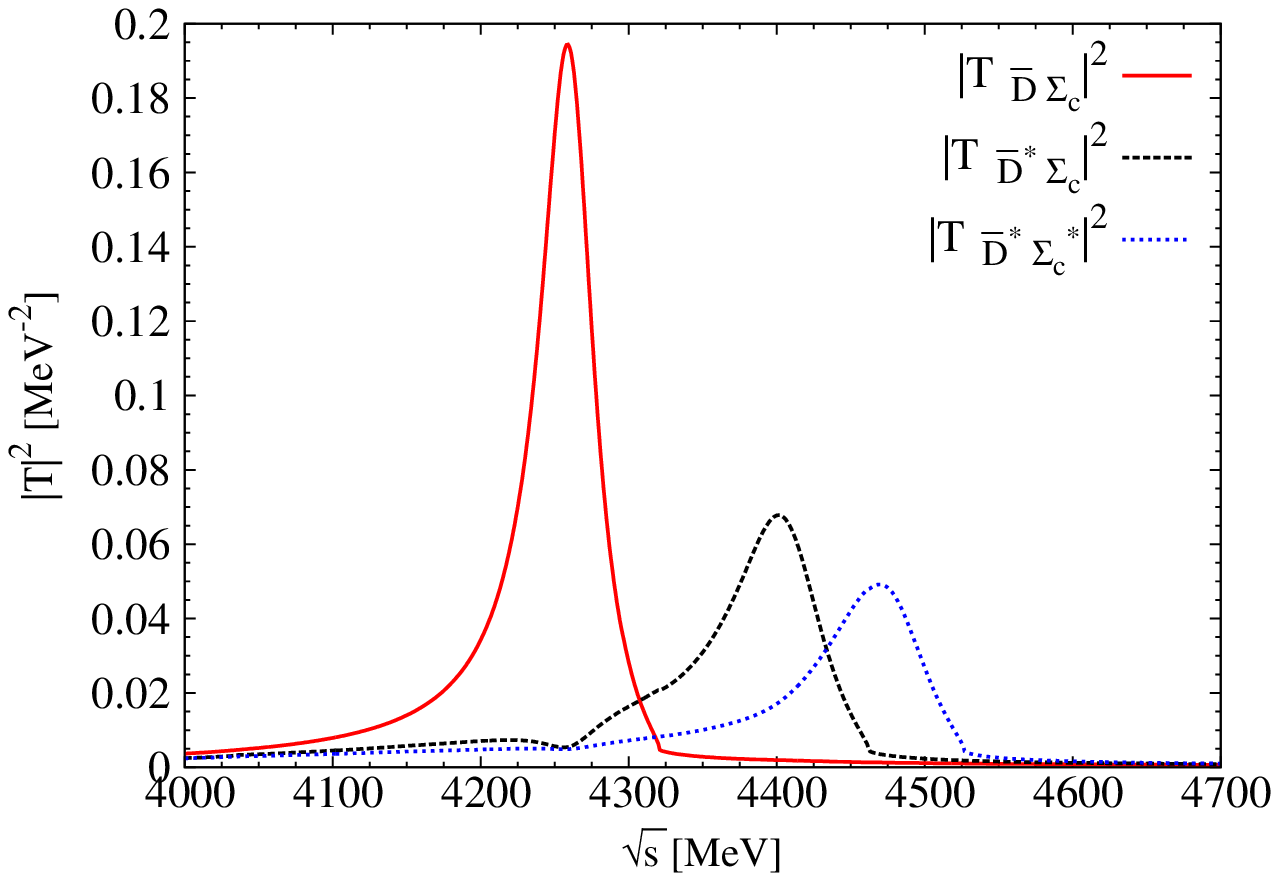}
\includegraphics[scale=0.6]{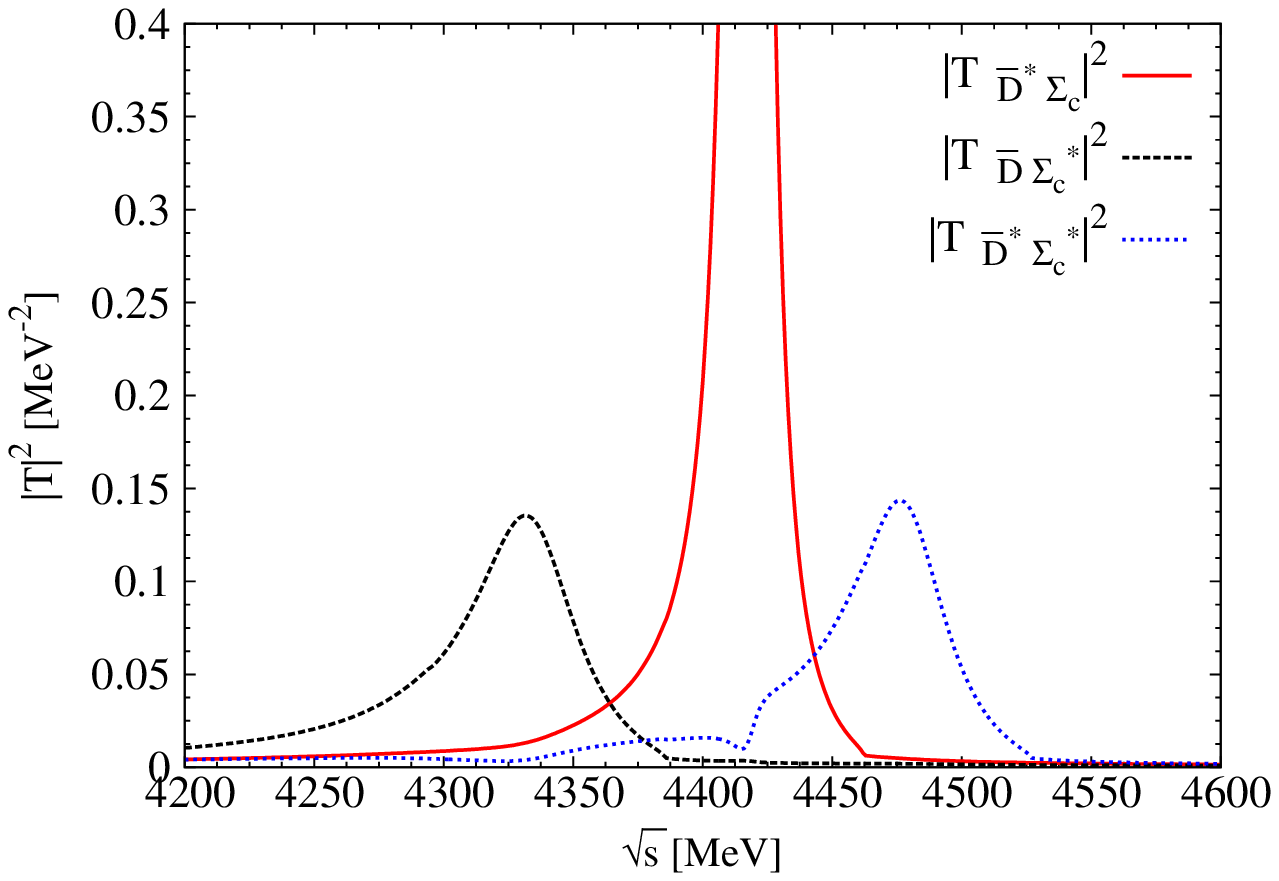}
\caption{Results of the modulus squared of the amplitudes. Left: $J=1/2,~I=1/2$ sector. Right: $J=3/2,~I=1/2$ sector.}
\label{fig:tsqam}
\end{figure}

We show our results of the modulus squared of the amplitudes in Fig.~\ref{fig:tsqam} for the $J=1/2,~I=1/2$ sector (left) and the $J=3/2,~I=1/2$ sector (right), which are consistent with the ones obtained in Refs. \cite{Xiao:2013yca,Xiao:2015fia}. For the $J=1/2,~I=1/2$ sector, corresponding to the peaks in the modulus squared of the amplitudes, we find the poles in the second Riemann sheets as $(4260.9 - i\;22.1)\mev$, $(4408.5 - i\;37.0)\mev$ and $(4479.0 - i\;39.9)\mev$, which are under the thresholds of the channels $\bar D \Sigma_c$, $\bar D^* \Sigma_c$ and $\bar D^* \Sigma^*_c$ respectively. For the $J=3/2,~I=1/2$ sector, we find the corresponding poles as $(4335.0 - i\;24.2)\mev$, $(4418.0 - i\;5.1)\mev$ and $(4479.2 - i\;21.7)\mev$, which are slightly below the thresholds of $\bar D \Sigma_c^*$, $\bar D^* \Sigma_c$ and $\bar D^* \Sigma^*_c$ channels, respectively. Compared to the results of Refs. \cite{Xiao:2013yca,Xiao:2015fia}, the masses of the poles are just a few MeV different, but the widths differ by $5 \sim 20 \mev$. These results are also consistent with the findings of Refs. \cite{Wu:2010jy,Wu:2010vk,Ortega:2016syt} within the uncertainties. 

\section{The s-/u- channel contributions}

\begin{figure}
\centering
\includegraphics[scale=0.6]{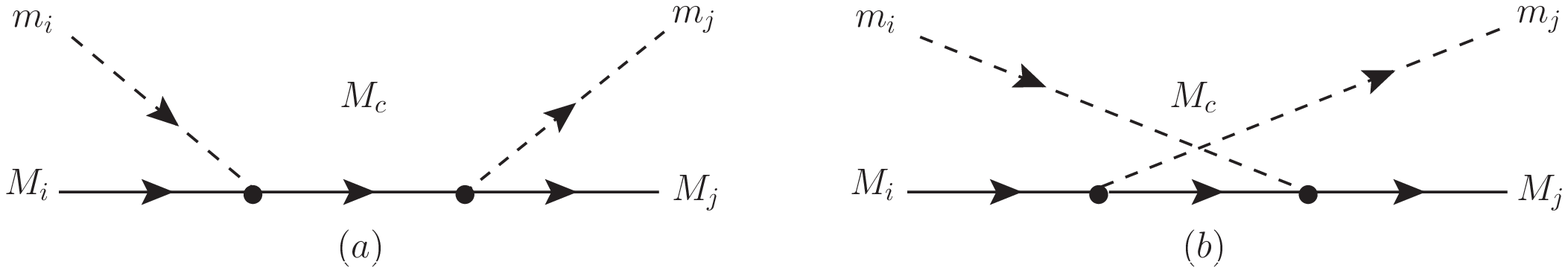}
\caption{The diagrams of the s-channel (a) and u-channel (b) contributions.}
\label{fig:sudia}
\end{figure}

Now we take into account the s-/u- channel contributions to our interaction potentials of Eq. \eqref{eq:vpoten} as done in Refs. \cite{Oller:2000fj,Borasoy:2005ie,Oller:2005ig,Borasoy:2006sr,Ikeda:2012au,Hyodo:2011ur,Ramos:2016odk}, shown in Fig. \ref{fig:sudia}. In fact, the potentials of Eq. \eqref{eq:vpoten} are the WT type called $V_{ij}^{WT}$, where we only consider the t-channel diagrams with the local hidden gauge formalism. More details about the WT type interactions have already discussed in Ref. \cite{Borasoy:1995ds} for the presence of vector mesons, the massive spin 1 fields. Thus, for the present case we have
\begin{equation}
V_{ij} = V_{ij}^{WT} + V_{ij}^s + V_{ij}^u \; .
\end{equation}

To evaluate the diagrams of Fig. \ref{fig:sudia}, we need to calculate the interaction vertices from the corresponding Lagrangians. For the PB interactions, the lowest order ($\mathcal{O} (p)$) chiral Lagrangian is given by \cite{Pich:1995bw,Bernard:1995dp}
\begin{equation}
\mathcal{L}^{(1)}_{MB} =\langle \bar{B}i\gamma^\mu \nabla_\mu B - m_{0} \bar{B}B +\frac{D}{2}(\bar{B}\gamma^{\mu}\gamma_{5} \{u_{\mu},B\}) +\frac{F}{2}(\bar{B}\gamma^{\mu}\gamma_{5} [u_{\mu},B]) \rangle\; ,
\label{eq:lag1}
\end{equation}
where $u$ is the chiral field constructed by the meson fields $P$, $u = \exp [i P / (\sqrt{2}f)]$, $B$ the baryon fields, and $m_0$ denotes the common (average) mass of the octet baryons in the chiral limit. Besides, the low energy constants $D$ and $F$ are related to the usual nucleon axial-vector coupling constant $g_A=D+F = 1.26$ \cite{Pich:1995bw,Bernard:1995dp}. By expanding the covariant derivative term, the Lagrangian of Eq. \eqref{eq:lag1} will lead to the contact term of WT type and the Born term of Yukawa vertex,
\begin{align}
\mathcal{L}_{PBPB}^{WT} &=\frac{1}{4f^2} \langle \bar{B}i\gamma^{\mu}[\,[P, \partial_{\mu}P],B] \rangle\; ; \label{eq:Lagwt}  \\
\mathcal{L}_{BBP} &=-\frac{1}{\sqrt{2}f} \langle D(\bar{B}\gamma^{\mu}\gamma_{5}\{\partial_{\mu}P,B\}) +F(\bar{B}\gamma^{\mu}\gamma_{5} [\partial_{\mu}P,B]) \rangle\; . \label{eq:Lagyu}
\end{align}
From the Lagrangian of Eq. \eqref{eq:Lagwt}, we can obtain the tree-level amplitude, written \footnote{Note that, since the phase convention and the normalization are different from Refs. \cite{Oller:2000fj,Borasoy:2005ie,Oller:2005ig,Borasoy:2006sr,Ikeda:2012au,Hyodo:2011ur,Mai:2012dt,Guo:2012vv,
Mai:2014xna,Cieply:2016jby,Ramos:2016odk}, there is a factor $-2\sqrt{M_i M_j}$ or $\sqrt{M_i M_j}$ between them,  and thus in the final expressions for the scattering amplitudes $T_{ij}$.}
\begin{align}
V_{i j}^{WT} (\sqrt{s},\Omega,\sigma_i,\sigma_j) =&-\frac{C_{i j}}{4\; f^2} \;  N_i\;N_j \; (\chi^{\sigma_j})^\dagger \Bigg[ 2\sqrt{s}-M_i-M_j \nonumber  \\
&+ (2\sqrt{s}+M_i+M_j) \frac{\vec{p}_i\cdot\vec{p}_j + i(\vec{p}_i\times\vec{p}_j)\cdot\vec{\sigma}} {(M_i+E_i)(M_j+E_j)} \Bigg] \chi^{\sigma_i}\; ,
\end{align}
where $\vec{p}_i$ is the three momentum of the baryon in corresponding channel $i$, $C_{ij}$ the coefficients, and $\chi^{\sigma_i}$ is the two-component Pauli spinor for the baryon in corresponding channel $i$. Then, we should do the s-wave projection with the solid angle $\Omega$ of the scattering and the spin summation of $\sigma_i$, using
\begin{equation}
V_{ij}^{WT}(\sqrt{s}) =\frac{1}{8\pi} \sum_{\sigma_i, \, \sigma_j}\int d\Omega \ V_{ij}(W,\Omega,\sigma_i,\sigma_j)\; .
\end{equation}
Thus, we obtain the WT interaction potential
\begin{equation}
V_{i j}^{WT}(\sqrt{s}) = - \frac{C_{i j}}{4 f^2} \;  N_i\;N_j\; (2\sqrt{s} - M_{i}-M_{j})\; ,
\end{equation}
which are indeed the types in Eq. \eqref{eq:vpoten} from the Local hidden gauge Lagrangian for the light vector meson exchange cases (WT type, see the later discussions in this section), but not for the heavy vector exchange cases, see $\mu_{12}$ in Eq. \eqref{eq:vpoten}.

Using the Lagrangian of Eq. \eqref{eq:Lagyu}, the vertex can be derived where one can substitute an $s$ quark by a $c$ quark for the corresponding hadrons in the SU(3) symmetry as done in Ref.  \cite{Xiao:2013yca}, having
\begin{equation}
V_{ie}^{BBP} = C_{i,e}^P\; \bar{u}_e(p_e,M_e)\; \gamma_\mu \gamma_5\; u_i(p_i,M_i)\; k^\mu \; ,
\end{equation} 
where $u_i(p_i,M_i)$ is the corresponding baryon spinor and $k$ the incoming momentum of the pseudoscalar meson. We only have the nucleon $N$ as propagating particle in our present case, thus, the coefficients $C_{ie}^P$ with isospin $I = 1/2$ are given by
\begin{equation}
C_{\Lambda_c,N}^{\bar{D}} = \frac{1}{\sqrt{3}} \frac{D+3F}{2f}, \qquad C_{\Sigma_c,N}^{\bar{D}} = \frac{3}{\sqrt{3}} \frac{D-F}{2f} \; .
\end{equation}
Then, we can obtain the potentials from the contributions of the s-/u- channels as shown in Fig. \ref{fig:sudia}, for the cases of $PB \to PB$ transitions \footnote{As a check, these expressions could also be obtained using a Mathematica package of FeynCalc \cite{Mertig:1990an}.},
\begin{align}
V^s_{ij}(\sqrt{s},\Omega,\sigma_i,\sigma_j) =& C_{i,N}^P \; C_{j,N}^P\; \frac{1}{s-M_N^2} \;N_i\; N_j \; (\chi^{\sigma_j})^\dagger \Bigg\{ (\sqrt{s}-M_N)\big[s-\sqrt{s}(M_i+M_j)+M_iM_j\big] \nonumber \\
 & +(\sqrt{s}+M_N)\big[s+\sqrt{s}(M_i+M_j)+M_iM_j\big] \frac{\vec{p}_i\cdot\vec{p}_j +i(\vec{p}_i\times\vec{p}_j)\cdot\vec{\sigma}}{(M_i+E_i)(M_j+E_j)}\Bigg\} \chi^{\sigma_i}\; ,  \\
V^u_{ij}(\sqrt{s},\Omega,\sigma_i,\sigma_j) =& -C_{i,N}^P \; C_{j,N}^P\; \frac{1}{u-M_N^2}  \;N_i\; N_j \; (\chi^{\sigma_j})^\dagger \Bigg\{ u(\sqrt{s}+M_N) + \sqrt{s}\big[M_iM_j+M_N(M_i \nonumber \\
& +M_j)\big]-M_iM_jM_N -M_i^2(M_j+M_N)-M_j^2(M_i+M_N)+ \big\{ u(\sqrt{s}-M_N)  \nonumber \\
&  +\sqrt{s}\big[M_iM_j+M_N(M_i+M_j)\big]+M_iM_jM_N +M_i^2(M_j+M_N)+M_j^2(M_i  \nonumber \\
& +M_N) \big\}\; \frac{\vec{p}_i\cdot\vec{p}_j +i(\vec{p}_i\times\vec{p}_j)\cdot\vec{\sigma}}{(M_i+E_i)(M_j+E_j)} \Bigg\} \chi^{\sigma_i} \; ,
\end{align}
where $u = -s+m_i^2 +m_j^2 +2E_i E_j-2\vec{p}_i\cdot\vec{p}_j$. After $s$-wave projection, we obtain
\begin{align}
V^s_{ij}(\sqrt{s}) =& C_{i,N}^P \; C_{j,N}^P  \;N_i\; N_j \; \frac{1}{\sqrt{s}+M_N}\; (\sqrt{s}-M_i)(\sqrt{s}-M_j)\; , \label{eq:vijs1}  \\
V^u_{ij}(\sqrt{s}) =& -C_{i,N}^P \; C_{j,N}^P \;N_i\; N_j \; \Bigg\{ \sqrt{s}+M_N -\frac{(M_i+M_N)(M_j+M_N)}{2(M_i+E_i)(M_j+E_j)} (\sqrt{s}-M_N+M_i+M_j) \nonumber \\
& +\frac{(M_i+M_N)(M_j+M_N)}{4q^{cm}_iq^{cm}_j} \Big[ \sqrt{s}+M_N-M_i-M_j -\frac{s+M_N^2-m_i^2-m_j^2-2E_iE_j}{2(M_i+E_i)(M_j+E_j)}\nonumber \\
& \times (\sqrt{s}-M_N+M_i+M_j)\Big] \ln\frac{s+M_N^2-m_i^2-m_j^2-2E_iE_j-2q^{cm}_iq^{cm}_j} {s+M_N^2-m_i^2-m_j^2-2E_iE_j+2q^{cm}_iq^{cm}_j} \Bigg\}\; .
\label{eq:viju1}
\end{align}

For the case of $PB \to VB$ transitions, we need to evaluate the $BBV$ vertex. The Lagrangian for the interactions of vector mesons to the baryons is given by \cite{Klingl:1997kf,Oset:2009vf}
\begin{equation}
\mathcal{L}_{BBV} =g\;(\langle\bar{B}\gamma_\mu[V^\mu,B]\rangle + \langle\bar{B}\gamma_\mu B\rangle \langle V^\mu\rangle)\; ,
\end{equation}
where the coupling $g=\frac{m_V}{2f}$, with the mass of the vector meson $m_V$, and which leads to
\begin{equation}
V_{ie}^{BBV} = i\;g\; C_{i,e}^V\; \bar{u}_e (p_e,M_e)\; \gamma^\mu\; u_i(p_i,M_i)\; \epsilon^{(\lambda)}_\mu \; ,
\end{equation}
with the polarization vector $\epsilon^{(\lambda)}$ of the incoming vector meson and the coefficients $C_{ie}^V$ with isospin $I = 1/2$ for our present case as
\begin{equation}
C_{\Lambda_c,N}^{\bar{D^*}} = -\frac{3}{\sqrt{6}}\; g, \qquad C_{\Sigma_c,N}^{\bar{D^*}} = -\frac{3}{\sqrt{6}}\; g \; .
\end{equation}
Thus, we can evaluate the potentials for the $PB \to VB$ transitions,
\begin{align}
V^s_{ij}(\sqrt{s},\Omega,\sigma_i,\sigma_j) =& - C_{i,N}^P \; C_{j,N}^V\; \frac{i}{s-M_N^2} \;N_i\; N_j \; (\chi^{\sigma_j})^\dagger \Bigg\{ s+M_iM_N-\sqrt{s}(M_i+M_N) \nonumber \\
 & +\big[s+M_iM_N+\sqrt{s}(M_i+M_N)\big] \frac{\vec{p}_i\cdot\vec{p}_j +i(\vec{p}_i\times\vec{p}_j)\cdot\vec{\sigma}}{(M_i+E_i)(M_j+E_j)}\Bigg\}\; \sigma^k \; \epsilon^{(\lambda_j)}_k \;\chi^{\sigma_i}\; ,  \\
V^u_{ij}(\sqrt{s},\Omega,\sigma_i,\sigma_j) =& C_{i,N}^P \; C_{j,N}^V\; \frac{i}{u-M_N^2}  \;N_i\; N_j \; (\chi^{\sigma_j})^\dagger \Bigg\{ \Bigg( \big[\sqrt{s}(M_j+M_N) - (M_iM_N+M_j^2 \nonumber \\
& +M_iM_j-u)\big]-\big[\sqrt{s}(M_j+M_N) + (M_iM_N+M_j^2+M_iM_j-u)\big]  \nonumber \\
& \times \frac{\vec{p}_i\cdot\vec{p}_j +i(\vec{p}_i\times\vec{p}_j)\cdot\vec{\sigma}}{(M_i+E_i)(M_j+E_j)} \Bigg)\; \sigma^k \; \epsilon^{(\lambda_j)}_k - 2 \vec{p}_i\cdot\vec{\epsilon}\;^{(\lambda_j)}\; (M_j+M_N)\; \sigma_k \; \Big(\frac{p_i^k}{M_i+E_i}  \nonumber \\
& -\frac{p_j^k}{M_j+E_j} \Big) \Bigg\} \;\chi^{\sigma_i} \; ,
\end{align}
where we only take into account the space-like polarization and ignore the unphysical time-like one as done in Refs. \cite{Khemchandani:2012ur,Kang:2016zmv}. But, after the spin summation, we found that
\begin{equation}
\sum_{\sigma_i,\;\sigma_j} \;(\chi^{\sigma_j})^\dagger \;\sigma_k\; \chi^{\sigma_i} \delta_{\sigma_i,\;\sigma_j} =0 \;,
\end{equation}
where $\delta_{\sigma_i,\;\sigma_j}$ is the usual Kronecker delta function, which is a consequence of $\sigma_k$ having spin operator properties. Therefore, for the $PB \to VB$ transitions, the s-/u- channel contributions are zero and consistent with the results of Ref. \cite{Cabrera:2000dx} where a vanishing contribution for the $\pi N \to \rho N$ transitions was found. In principle, as discussed in Ref. \cite{Garzon:2012np}, there are also contributions from the meson in flight term and the contact term (Kroll-Ruderman term) for the $PB \to VB$ transitions which can be obtained by introducing $PB$ and $VB$ interaction Lagrangian as done in Ref. \cite{Khemchandani:2011mf}. In fact, the meson in flight terms are concerned in the $J/\psi N$ and $\eta_c N$ related channels for the heavy meson exchange. But, for the others channels, we also have checked that the contributions from these two terms are trivial as discussed in Ref. \cite{Xiao:2013yca}, and thus, we ignore them.

Then, for the $VB \to VB$ transitions, the amplitudes of the s-/u- channel diagrams are given by
\begin{align}
V^s_{ij}(\sqrt{s},\Omega,\sigma_i,\sigma_j) =& C_{i,N}^V \; C_{j,N}^V\; \frac{1}{s-M_N^2} \;N_i\; N_j \; (\chi^{\sigma_j})^\dagger \Bigg[ \sqrt{s}-M_N +\big(\sqrt{s}+M_N\big) \nonumber \\
 & \times \frac{\vec{p}_i\cdot\vec{p}_j +i(\vec{p}_i\times\vec{p}_j)\cdot\vec{\sigma}}{(M_i+E_i)(M_j+E_j)}\Bigg]\; \sigma^k \; \epsilon^{(\lambda_i)}_k \; \sigma^l \; \epsilon^{(\lambda_j)}_l \;\chi^{\sigma_i}\; ,  \\
V^u_{ij}(\sqrt{s},\Omega,\sigma_i,\sigma_j) =& -C_{i,N}^V \; C_{j,N}^V\; \frac{1}{u-M_N^2}  \;N_i\; N_j \; (\chi^{\sigma_j})^\dagger \Bigg\{\Bigg[ \sqrt{s}+M_N-M_i-M_j +\big(\sqrt{s}-M_N \nonumber \\
 & +M_i+M_j\big) \frac{\vec{p}_i\cdot\vec{p}_j +i(\vec{p}_i\times\vec{p}_j)\cdot\vec{\sigma}}{(M_i+E_i)(M_j+E_j)}\Bigg]\; \sigma^k \; \epsilon^{(\lambda_i)}_k \; \sigma^l \; \epsilon^{(\lambda_j)}_l +2\;\vec{p}_i\cdot\vec{\epsilon_j}\; \vec{\sigma}\cdot\vec{\epsilon_i}\;\sigma_k \nonumber \\
& \times  \Big(\frac{p_i^k}{M_i+E_i}+\frac{p_j^k}{M_j+E_j} \Big) +2\;\vec{p}_j\cdot\vec{\epsilon_i}\; \vec{\sigma}\cdot\vec{\epsilon_j}\;\sigma_k \; \Big(\frac{p_i^k}{M_i+E_i}+\frac{p_j^k}{M_j+E_j} \Big)  \Bigg\}\;\chi^{\sigma_i} \; .
\end{align}
Then, doing $s$-wave projection, we find
\begin{align}
V^s_{ij}(\sqrt{s}) =& C_{i,N}^V \; C_{j,N}^V  \;N_i\; N_j \; \frac{1}{\sqrt{s}+M_N}\; \vec{\sigma}\cdot\vec{\epsilon_i} \;\vec{\sigma}\cdot\vec{\epsilon_j} \; , \label{eq:vijs2}  \\
V^u_{ij}(\sqrt{s}) =& -C_{i,N}^V \; C_{j,N}^V \;N_i\; N_j \; \Bigg\{-\frac{\sqrt{s}-M_N+M_i+M_j}{2(M_i+E_i)(M_j+E_j)} +\frac{1}{4q^{cm}_iq^{cm}_j} \Big[ \sqrt{s}+M_N-M_i-M_j  \nonumber \\
&  -\frac{s+M_N^2-m_i^2-m_j^2-2E_iE_j}{2(M_i+E_i)(M_j+E_j)} (\sqrt{s}-M_N+M_i+M_j)\Big] \nonumber \\
& \times \ln\frac{s+M_N^2-m_i^2-m_j^2-2E_iE_j-2q^{cm}_iq^{cm}_j} {s+M_N^2-m_i^2-m_j^2-2E_iE_j+2q^{cm}_iq^{cm}_j} \Bigg\}\; \vec{\sigma}\cdot\vec{\epsilon_i} \;\vec{\sigma}\cdot\vec{\epsilon_j}\; ,
\label{eq:viju2}
\end{align}
where we also have the structure $\vec{\sigma}\cdot\vec{\epsilon_i} \;\vec{\sigma}\cdot\vec{\epsilon_j}$ as well as the ones obtained in Ref. \cite{Khemchandani:2012ur}.

We first add the contributions from s-channel diagram and find that the modulus squared of the amplitudes are not much different from the ones in Fig. \ref{fig:tsqam} and the peaks move just a bit. Thus, the corresponding poles are changed to $(4264.0 - i\;23.2)\mev$, $(4411.2 - i\;38.0)\mev$ and $(4474.9 - i\;40.9)\mev$ below the thresholds of $\bar D \Sigma_c$, $\bar D^* \Sigma_c$ and $\bar D^* \Sigma^*_c$ channels respectively in the $J=1/2,~I=1/2$ sector, and $(4338.4 - i\;26.2)\mev$, $(4420.8 - i\;5.9)\mev$ and $(4478.9 - i\;24.1)\mev$ below the thresholds of $\bar D \Sigma_c^*$, $\bar D^* \Sigma_c$, $\bar D^* \Sigma^*_c$ channels respectively in the $J=3/2,~I=1/2$ sector. These results are just a few MeV different both in the masses and the widths of the poles compared to the ones that we have in the last section. Indeed, for the results we have here, the contributions from s-channel diagram is suppressed by the nucleon propagator, seen Eqs. \eqref{eq:vijs1} and \eqref{eq:vijs2}, and thus, the results with its contributions are not affected much.

\begin{figure}
\centering
\includegraphics[scale=0.6]{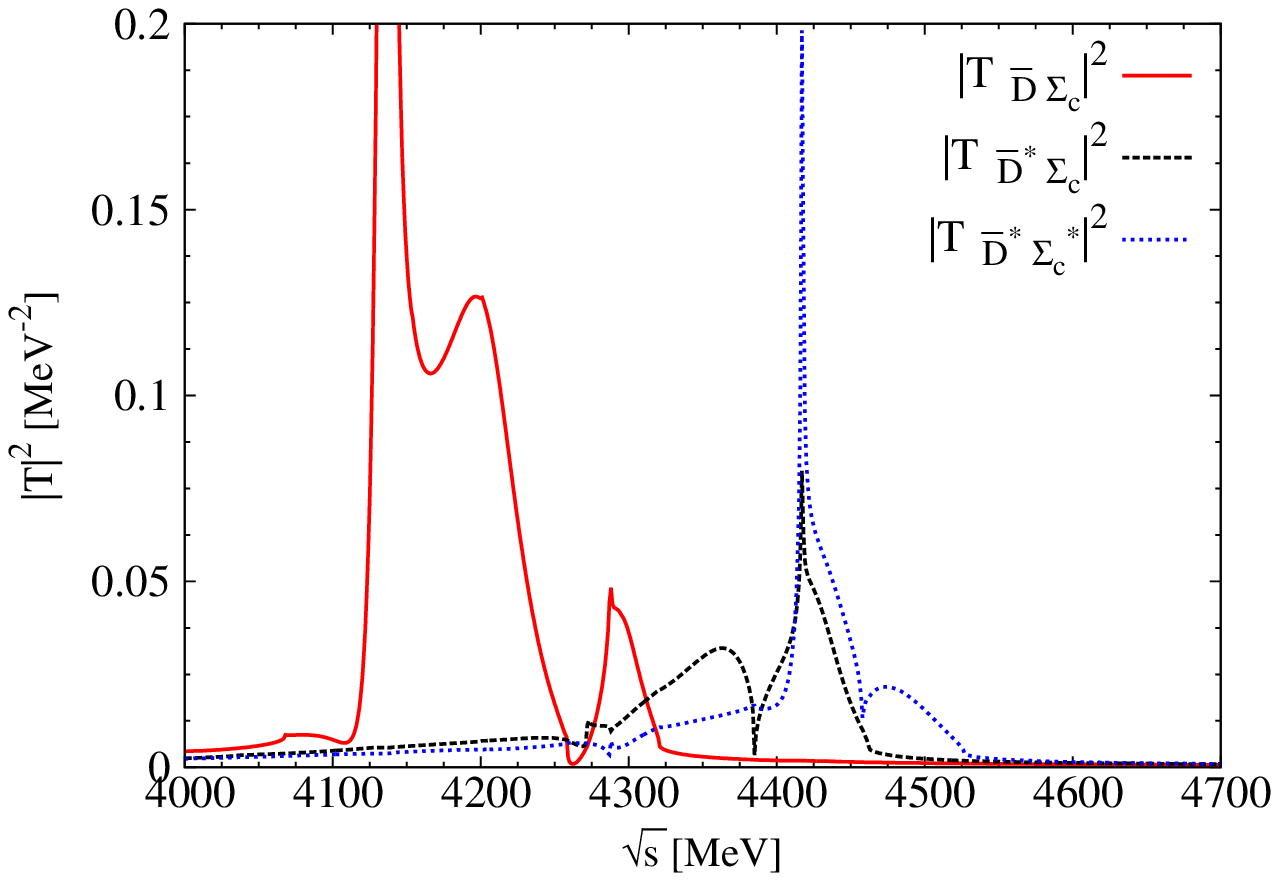}
\includegraphics[scale=0.6]{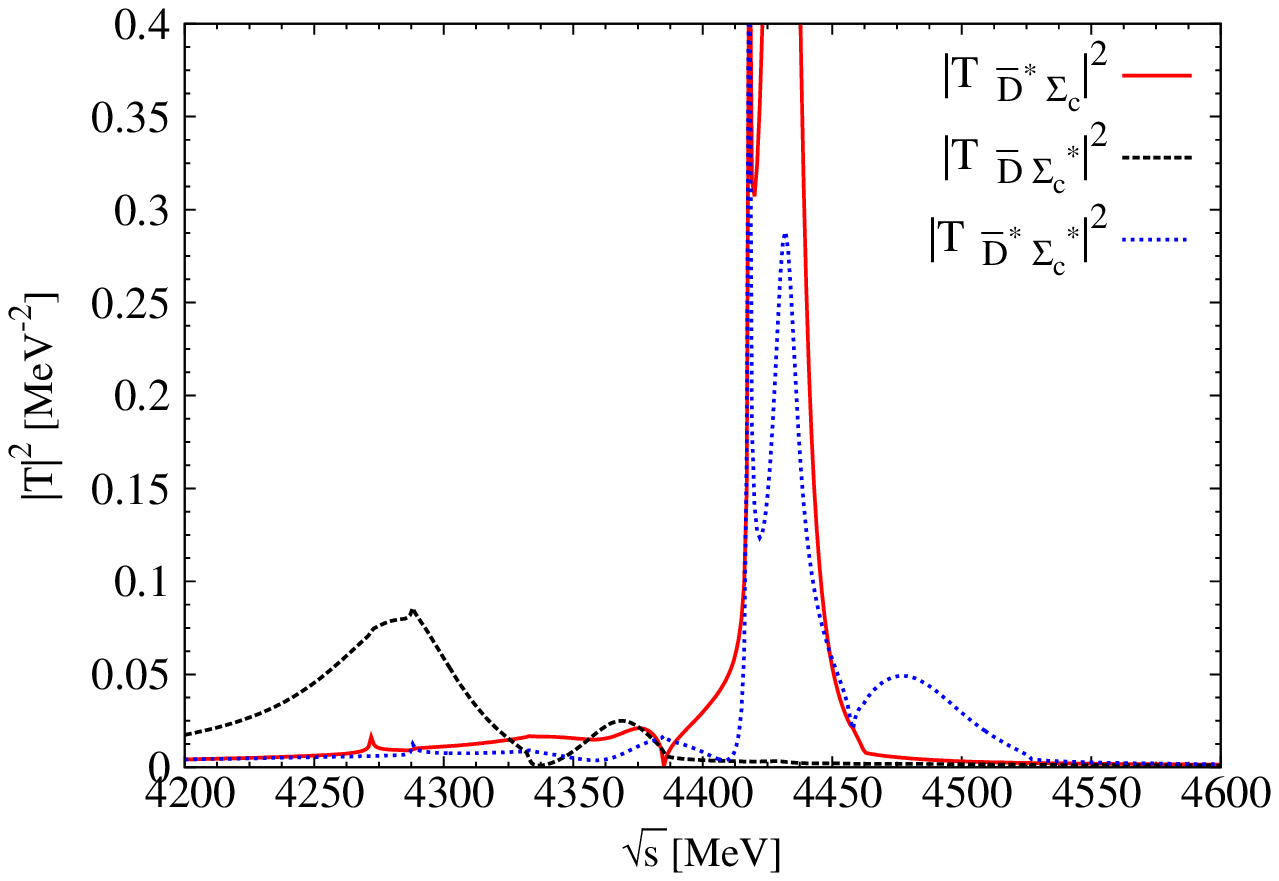}
\caption{Results of the modulus squared of the amplitudes with all the contributions from s-/u- channels. Left: $J=1/2,~I=1/2$ sector. Right: $J=3/2,~I=1/2$ sector.}
\label{fig:tsqam2}
\end{figure}

Then we add all the contributions from the s-/u- channel diagrams. The results are shown in Fig. \ref{fig:tsqam2}, where we can see a mess in the modulus squared of the amplitudes and not so clear peaks as in Fig. \ref{fig:tsqam} since we have added the u-channel contributions. We find that these extra unphysical (sharp) peaks come from the contributions of the unphysical subthreshold effects in the u-channel as discussed in Ref. \cite{Borasoy:2005ie} which appear at some certain energies. But, in the case of $\bar{K} N$ (PB) interactions in the isospin $I=0$ sector, these unphysical effects are numerically small \footnote{Note that this does not hold for the $\bar{K} N$ interactions in the isospin $I=1$ sector, where important contributions of u-channel were found in Ref. \cite{Guo:2012vv} and lead to a pole most likely in the near-threhold region of $\bar{K} N$. Some technical details about the u-channel contributions in the isospin $I=1$ sector are discussed in Ref. \cite{Oller:2006ss}.} as discussed in Ref. \cite{Borasoy:2005ie}, and analogously ignored contributions of these diagrams were found in Ref. \cite{Sarkar:2009kx} in the case of VB interactions for the light quark sectors. Note that, as discussed at the beginning of this section, the WT type interaction potentials from the local hidden gauge formalism are in fact the t-channel diagrams with the explicit exchange of a vector meson, thus, in principle, there are also the unphysical left hand cut contributions in the t-channel as in the u-channel. The fact is that we have taken an approximation where the transferred momentum $q^2$ has been neglected in comparison with the masses of the exchanged vector mesons $m_V^2$. Therefore, there is no singularities appearing in the interaction potentials \footnote{More discussions can be referred to Sec. 2.2 of Ref. \cite{Sarkar:2009kx}.}, and this leads to the t-channel diagrams contributed as WT type for the light vector meson exchanges. Even though, these unphysical contributions from the u-channel do not affecte the pole in the second Riemann sheets since these unphysical effects only happen in the first Riemann sheet. We can still find clean poles in the second Riemann sheets and no unphysical poles, as $(4291.4 - i\;19.5)\mev$, $(4426.9 - i\;30.2)\mev$ and $(4482.4 - i\;61.3)\mev$ below the thresholds of $\bar D \Sigma_c$, $\bar D^* \Sigma_c$ and $\bar D^* \Sigma^*_c$ channels respectively in the $J=1/2,~I=1/2$ sector, and $(4367.6 - i\;21.0)\mev$, $(4432.5 - i\;5.4)\mev$ and $(4479.6 - i\;35.3)\mev$ below the thresholds of $\bar D \Sigma_c^*$, $\bar D^* \Sigma_c$, $\bar D^* \Sigma^*_c$ channels respectively in the $J=3/2,~I=1/2$ sector. We can see that there are up to 30 MeV differences in the masses and the width for some states compared to the ones with only s-channel contributions. It means that the contributions from u-channel are bigger than from s-channel. 

Finally, to summarize our results, we show the predicted states in Table. \ref{tab:sum}. From this table, we can see that the contributions from the s-channel are constructive interference effects and thus the masses and the widths of the predicted states increase a few MeV. But, as seen from the minus sign in Eqs. \eqref{eq:viju1} and \eqref{eq:viju2}, the contributions from the u-channel are destructive interference effects. Therefore, the widths of the states decrease except for the $\bar D^* \Sigma^*_c$ state which increases twice compared to the ones obtained in Refs. \cite{Xiao:2013yca,Xiao:2015fia}. These effects also lead to the increased masses of the states since the predicted states are bound states coming from the attractive interaction potentials. Especially, the $\bar D \Sigma_c$ and $\bar D \Sigma_c^*$ states less bound by nearly 30 MeV. In the experimental results \cite{Aaij:2015tga}, the width of $P_c(4380)$ state is about 205 MeV, and about 39 MeV for the $P_c(4450)$ state. For our results from the interaction in the free space, it is hard to distinguish them since there are some theoretical uncertainties discussed in Ref. \cite{Xiao:2015fia}. Therefore, in the next section we continue to investigate the $\Lambda_b^0$ decays where the two $P_c$ states are found in the LHCb experiments .

\begin{table}[htb]
\renewcommand{\arraystretch}{1.7}
     \setlength{\tabcolsep}{0.2cm}
\centering
\caption{The summarized results for the predicted states (units: MeV), shown in the format (mass, width).}
\label{tab:sum}
\begin{tabular}{lccc cccc}
\hline
sectors & Channels & Thresholds & Refs. \cite{Xiao:2013yca,Xiao:2015fia}  & WT term  & WT + s & WT + s + u &  Experiments   \\
\hline\hline
 & $\bar D \Sigma_c$     & 4320.8 & $(4262, ~35)$ & $(4261, ~44)$ & $(4264, ~46)$ & $(4291, ~39)$ & $\cdots \cdots$  \\
$J^P = \frac{1}{2}^-$
 & $\bar D^* \Sigma_c$   & 4462.2 & $(4410, ~58)$ & $(4409, ~74)$ & $(4411, ~76)$ & $(4427, ~60)$ & $P_c(4450)^+ ~?$  \\
 & $\bar D^* \Sigma^*_c$ & 4526.7 & $(4481, ~57)$ & $(4479, ~80)$ & $(4475, ~82)$ & $(4482, ~123)$ & $P_c(4450)^+ ~?$  \\
\hline
 & $\bar D \Sigma_c^*$   & 4385.3 & $(4334, ~38)$ & $(4335, ~48)$ & $(4338, ~52)$ & $(4368, ~42)$ & $P_c(4380)^+ ~?$  \\
$J^P = \frac{3}{2}^-$
 & $\bar D^* \Sigma_c$   & 4462.2 & $(4417, ~8) $ & $(4418, ~10)$ & $(4421, ~12)$ & $(4432, ~11)$ & $P_c(4450)^+ ~?$  \\
 & $\bar D^* \Sigma^*_c$ & 4526.7 & $(4481, ~35)$ & $(4479, ~44)$ & $(4479, ~48)$ & $(4480, ~71)$ & $P_c(4450)^+ ~?$   \\
\hline\hline
\end{tabular}
\end{table}

\section{$\Lambda_b^0 \to J/\psi K^- (\pi^-) p$ decays}

\begin{figure}
\centering
\includegraphics[scale=0.7]{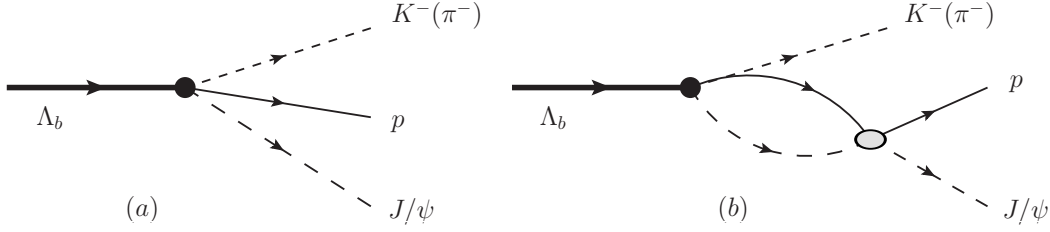}
\caption{The diagrams for the $\Lambda_b^0 \to J/\psi K^- (\pi^-) p$ decays: (a) direct $J/\psi K^- (\pi^-) p$ vertex decays at tree level; (b) final state interactions of $J/\psi p$.}
\label{fig:lambdec}
\end{figure}

We have calculated the interaction amplitudes of $J/\psi N$ and its coupled channels above, thus, we can go further to study the $J/\psi N$ final state interactions in the $\Lambda_b^0 \to J/\psi K^- (\pi^-) p$ decays where the two $P_c$ states are found in the $J/\psi p$ invariant mass distribution. Following the method used in Refs. \cite{Roca:2015dva,Roca:2015tea} (more details and discussions about this method can be found in Ref. \cite{Oset:2016nvf}, and references therein), as shown in Fig. \ref{fig:lambdec}, since the first diagram for the direct decays (a) only contributes a constant to the scattering amplitude, we only focus on the second diagram (b) of the $J/\psi N$ final state interactions. Thus, treating $K^-$ as spectator, for the $J/\psi p$ invariant mass distribution in the $\Lambda_b^0 \to J/\psi K^- p$ decays we have analogously
\begin{equation}
\frac{d \Gamma (M_{inv})}{d M_{inv}} = \frac{1}{4 (2\pi)^3} \; \frac{1}{M_{\Lambda_b}} \;\tilde{p}_{J/\psi}\; p_K |T^{J/\psi p}|^2 \; ,
\end{equation}
where $M_{inv}$ is the invariant mass of the $J/\psi p$ system in the final states, there is no $4 M_p$ factor for our normalization of the baryon spinors, the momenta in the center of mass frame are given by
\begin{align}
\tilde{p}_{J/\psi} (M_{inv}) &= \frac{\lambda^{1/2} (M_{inv}^2,m_{J/\psi}^2,M_p^2)}{2M_{inv}} \; ,  \\
p_K (M_{inv}) &= \frac{\lambda^{1/2} (M_{\Lambda_b}^2,m_K^2,M_{inv}^2)}{2M_{\Lambda_b}} \; ,
\end{align}
with $\lambda(a,b,c)=a^2+b^2+c^2-2(ab+ac+bc)$ the usual K\"all\'en function, and the transition amplitude of $T^{J/\psi p}$ is given by \cite{Wang:2015pcn}
\begin{equation}
T^{J/\psi p} (M_{inv}) = V_p\; h_{K^- p}\; G_{J/\psi p}(M_{inv}^2)\; t_{J/\psi p \to J/\psi p}(M_{inv}),
\end{equation}
with $h_{K^- p}=1$ \cite{Roca:2015tea}, the loop function $G_{J/\psi p}(M_{inv}^2)$ is given by Eq. \eqref{eq:giidr}, $V_p$ is a constant which collects the CKM matrix elements and the kinematic prefactors, and we take the amplitude of $t_{J/\psi p \to J/\psi p}(M_{inv})$ as $T^{I=1/2}_{J/\psi N}$ that we have evaluated in the former section and it is different from the one used in Refs. \cite{Roca:2015dva,Wang:2015pcn} by an approximation of the Breit-Wigner form. For the case of the $\Lambda_b^0 \to J/\psi \pi^- p$ decays, the formalism is analogous and just replaces $m_K$ with $m_\pi$.

\begin{figure}
\centering
\includegraphics[width=0.49\textwidth]{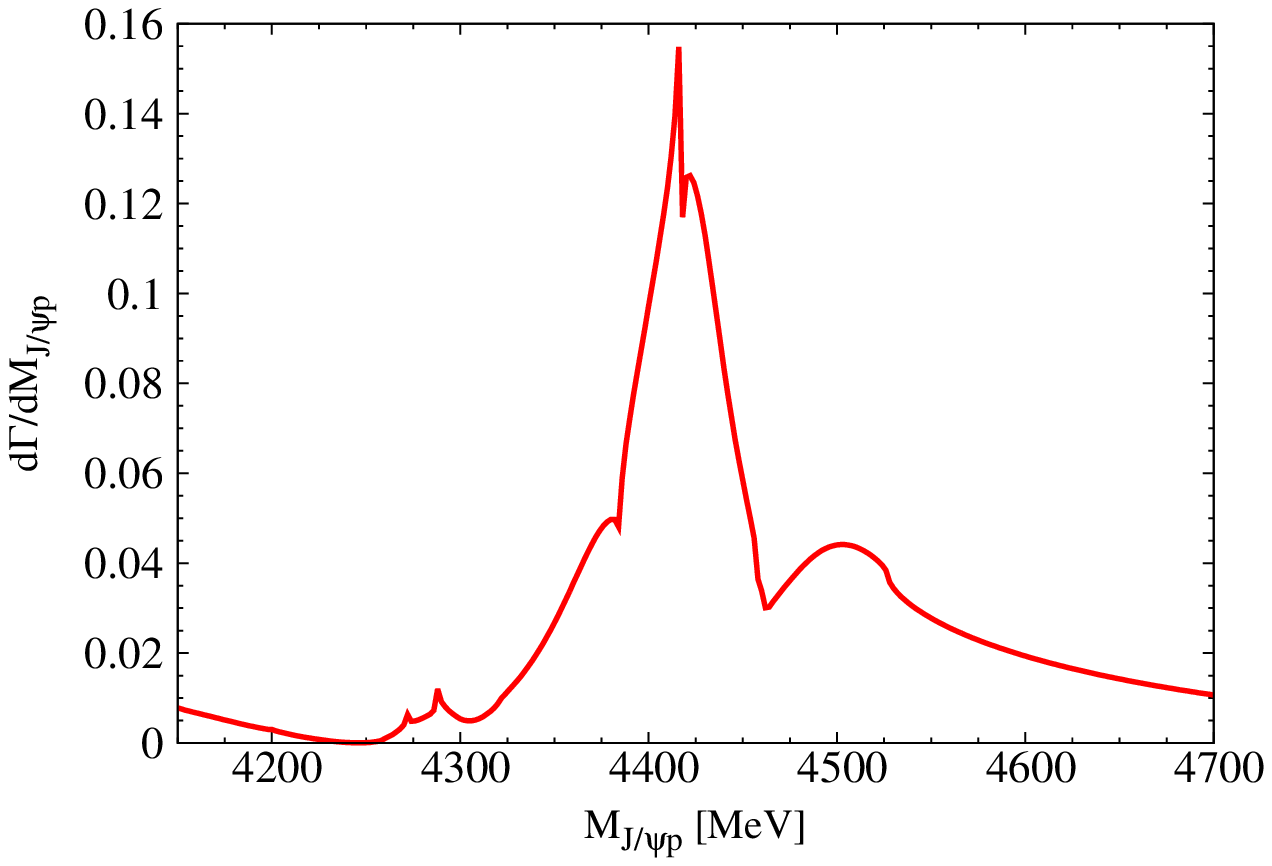}
\includegraphics[width=0.49\textwidth]{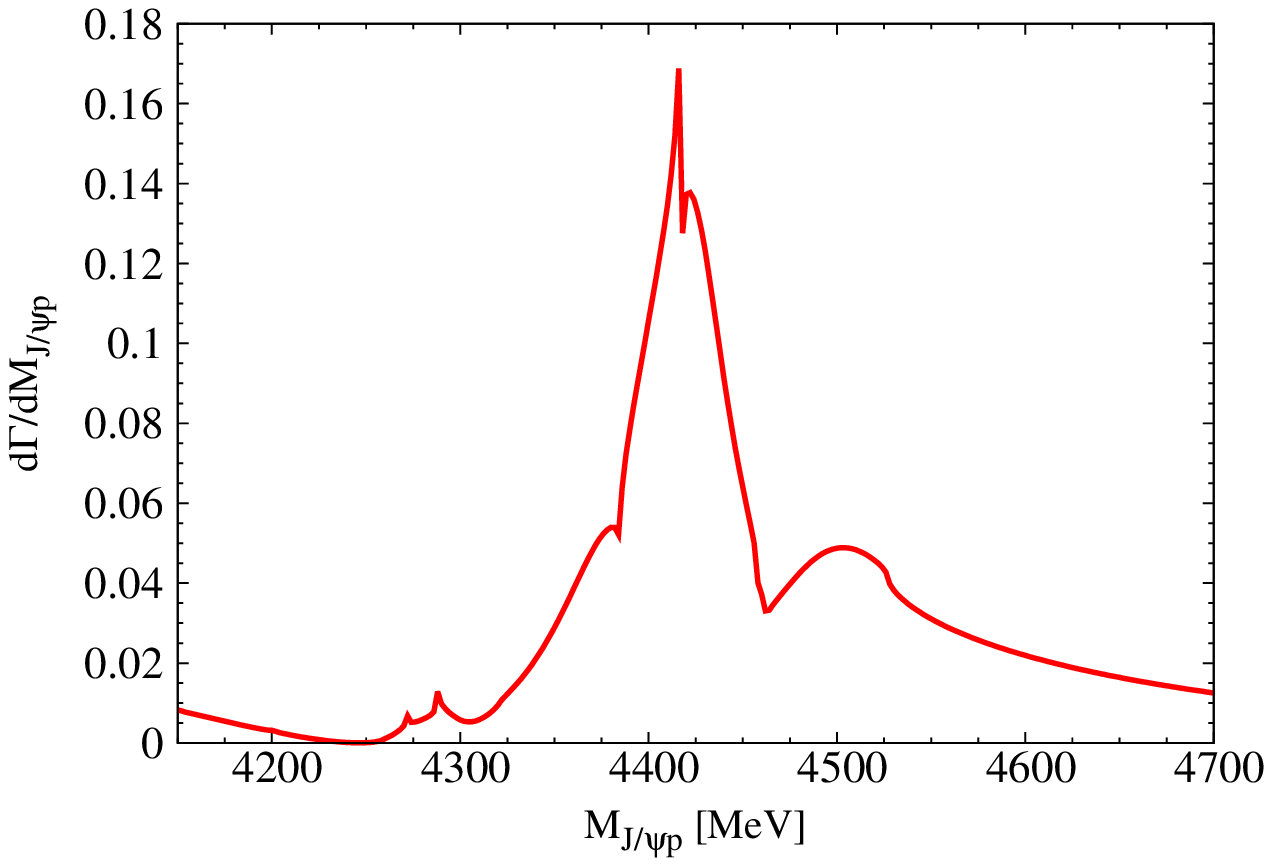}
\includegraphics[width=0.49\textwidth]{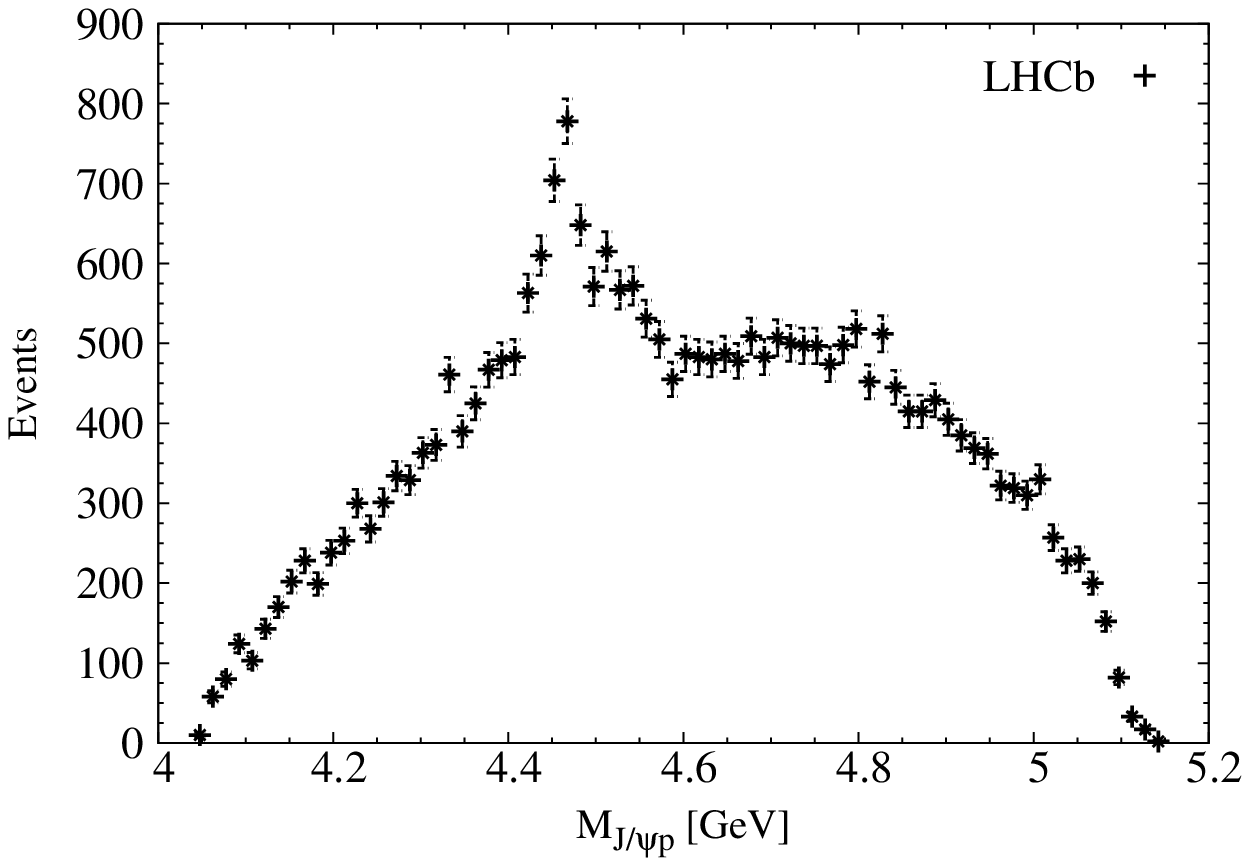}
\includegraphics[width=0.49\textwidth]{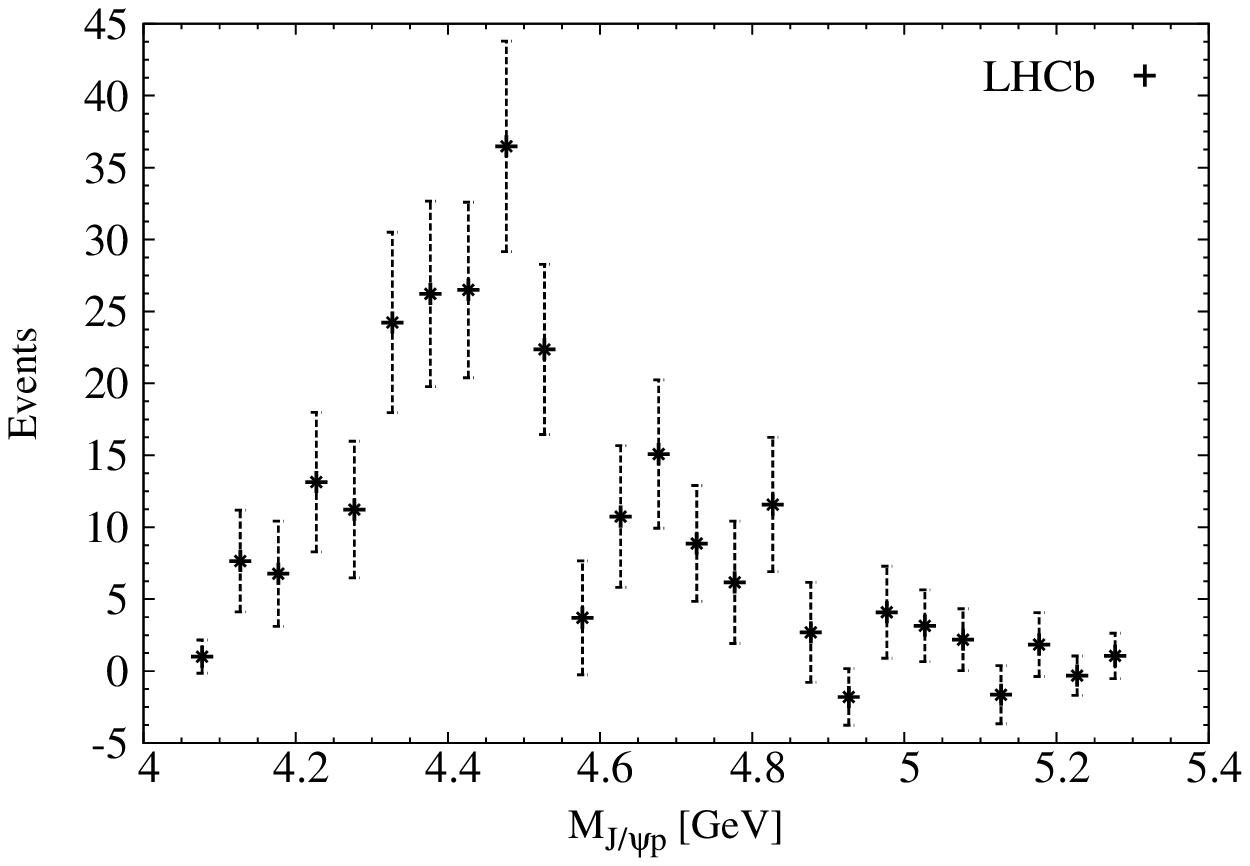}
\caption{Results of the $J/\psi p$ invariant mass distribution in $J=1/2$ sector: Upper-left panel for the $\Lambda_b^0 \to J/\psi K^- p$ decays; Upper-right for the $\Lambda_b^0 \to J/\psi \pi^- p$ decays. The experimental results are shown in the lower panels accordingly \cite{Aaij:2015tga,Aaij:2016ymb}.}
\label{fig:tsdga1}
\end{figure}

\begin{figure}
\centering
\includegraphics[width=0.49\textwidth]{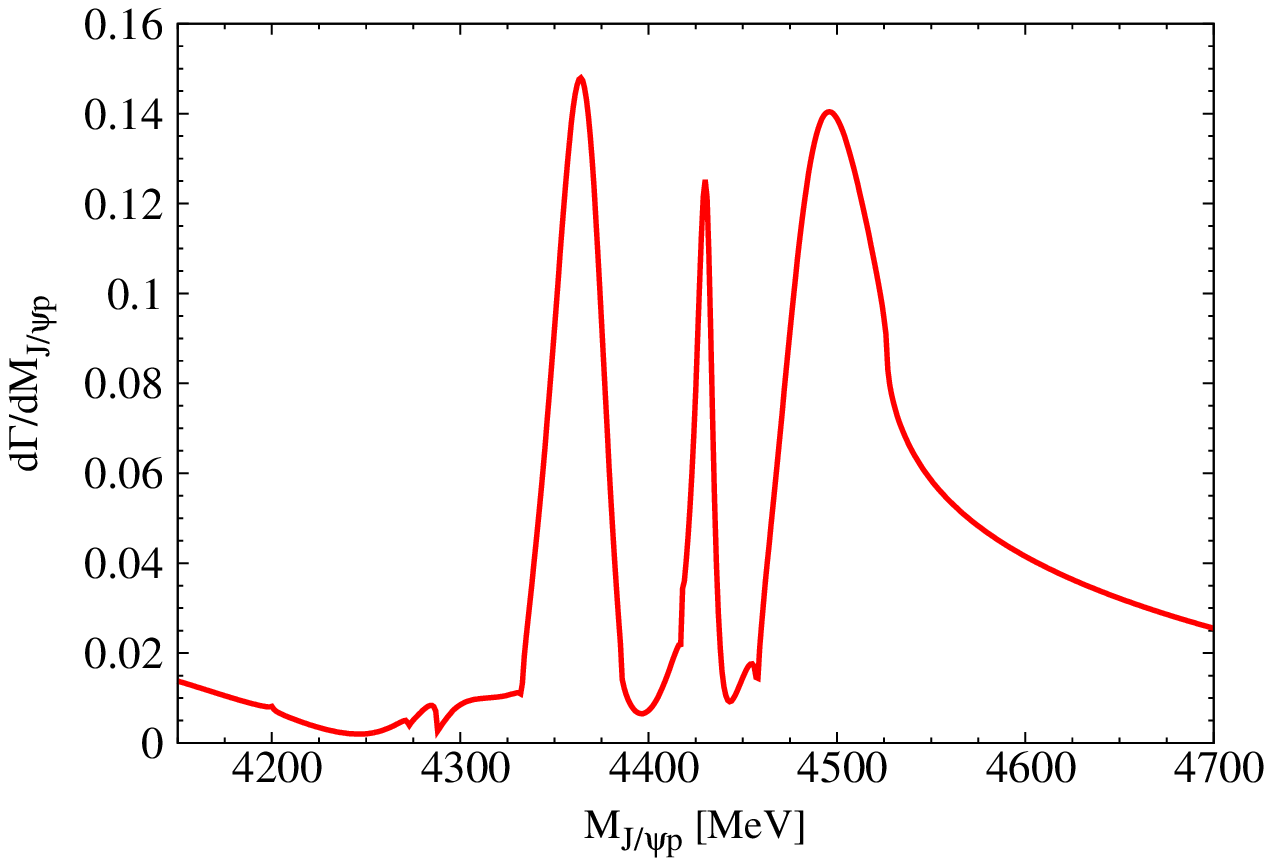}
\includegraphics[width=0.49\textwidth]{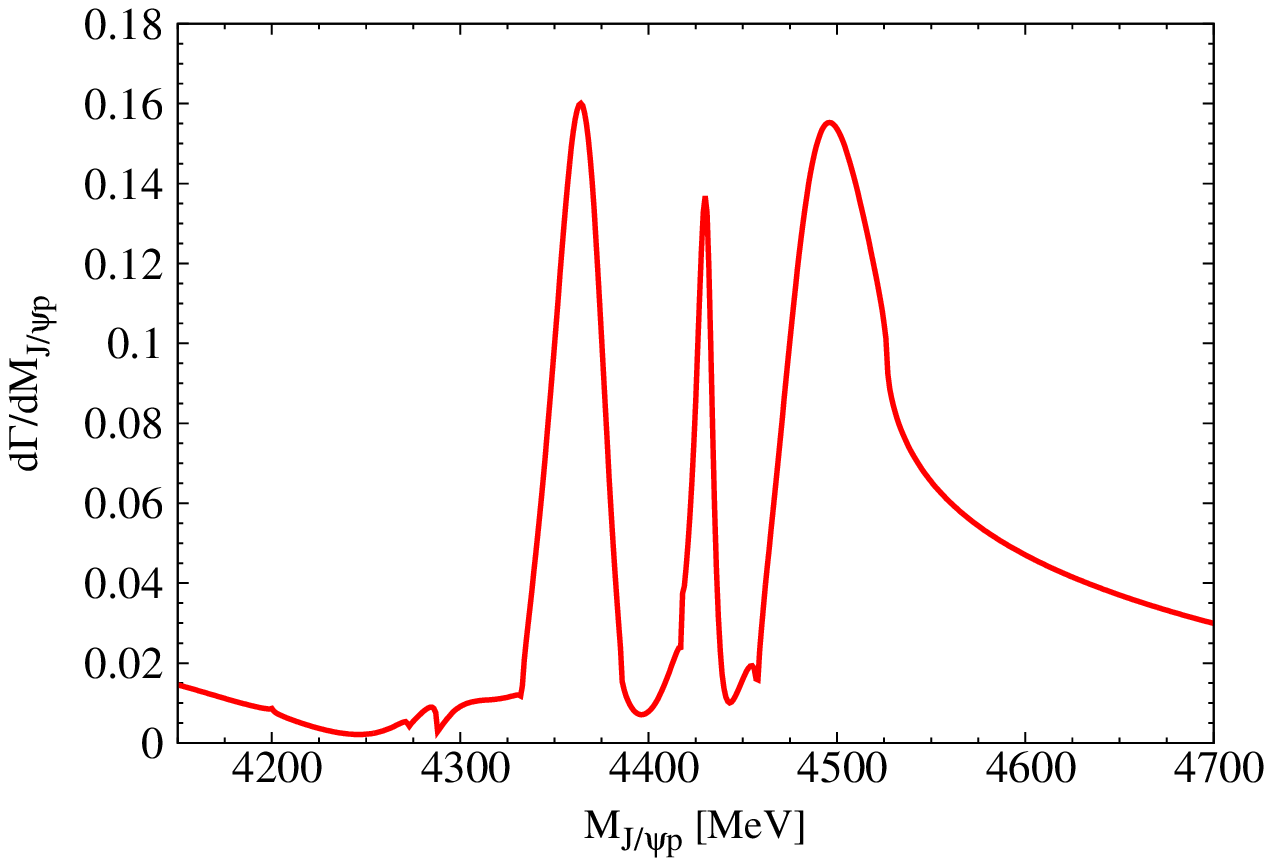}
\caption{Results of the $J/\psi p$ invariant mass distribution in $J=3/2$ sector: Left panel for the $\Lambda_b^0 \to J/\psi K^- p$ decays; Right for the $\Lambda_b^0 \to J/\psi \pi^- p$ decays.}
\label{fig:tsdga2}
\end{figure}

In our formalism in the former section, the scattering amplitudes for $T^{I=1/2}_{J/\psi N}$ have two cases, spins $J=1/2$ and $J=3/2$. Thus, using $J=1/2$ sector scattering amplitudes, we obtain the results as shown in Fig. \ref{fig:tsdga1} both for the $\Lambda_b^0 \to J/\psi K^- p$ (upper-left panel) and the $\Lambda_b^0 \to J/\psi \pi^- p$ (upper-right panel) decays where we have successfully produced the experimental line-shape found in the LHCb \cite{Aaij:2015tga,Aaij:2016phn,Aaij:2016ymb} and do not take into account the contributions of the background, even though there are three clear resonant peaks in the scattering amplitudes of free space, seen in the left of Figs. \ref{fig:tsqam} and \ref{fig:tsqam2}. From these results, we can see the clear resonant peak of $P_c (4450)$ state, and the structure of $P_c (4380)$ seems to appear. Therefore, from our formalism we can conclude that $P_c (4450)$ state could be a $J=1/2^-$ $\bar{D}^* \Sigma_c$ bound state. From the similar results with the $\Lambda_b^0 \to J/\psi K^- p$ and the $\Lambda_b^0 \to J/\psi \pi^- p$ decays, seen in the upper panels of Fig. \ref{fig:tsdga1}, indeed, the two $P_c$ states should be also seen in $\Lambda_b^0 \to J/\psi \pi^- p$ decays as suggested in the early research \cite{Burns:2015dwa,Wang:2015pcn} and found in Ref. \cite{Aaij:2016ymb}. Note that, we do not fit the experimental data as done in Ref. \cite{Wang:2015pcn}, since we do not take into account any background contributions and not change the parameters used in our former works. One should keep in mind that there are some uncertainties in our formalism as discussed above. For $J=3/2$ sector, the results are shown in Fig. \ref{fig:tsdga2} which can not match the experimental results for the $J/\psi p$ invariant mass distribution which have three resonant peaks of the same magnitude. These results also show that the missing two states in Fig. \ref{fig:tsdga1} for $J=1/2$ sector couple weakly with the $J/\psi N$ channel as found in Ref. \cite{Xiao:2013yca} and three predicted states having the same magnitude in Fig. \ref{fig:tsdga2} for $J=3/2$ sector couple strongly with the only decay channel of $J/\psi N$.

\section{Conclusions}

Using the local hidden gauge Lagrangian, combined with the heavy quark spin symmetry and the chiral symmetry, we re-investigated the interactions of $J/\psi N$ and its coupled channels, where we also took into account the contributions of the s-/u- channel diagrams ignored in the former works. As in the former works, the $\bar{D} \Sigma_c$, $\bar{D}^* \Sigma_c$, $\bar{D} \Sigma_c^*$ and $\bar{D}^* \Sigma_c^*$ bound states are found in the present work of which the masses and the widths are consistent with the former works within the uncertainties. Furthermore, we studied the $\Lambda_b^0 \to J/\psi K^- p$ and the $\Lambda_b^0 \to J/\psi \pi^- p$ decays with the scattering amplitudes of $J/\psi N$ obtained, we find consistent results with the LHCb experiments  \cite{Aaij:2015tga,Aaij:2016phn,Aaij:2016ymb} where we can associate the $P_c (4450)$ state as a $\bar{D}^* \Sigma_c$ bound state with $J=1/2^-$ and the structure of $P_c (4380)$ seems to appear too.

\section*{Acknowledgments}

The author thanks Ulf-G. Mei{\ss}ner, J. Gegelia and J. A. Oller for the useful comments and discussions, and acknowledges E. Wang for valuable comments. The experimental data information from S. L. Stone, L. Zhang, X. Song are appreciated.
This work is supported in part by the DFG and the NSFC through
funds provided to the Sino-German CRC~110 ``Symmetries and
the Emergence of Structure in QCD''.

\end{document}